\documentclass[amsfonts,amsmath,amssymb,showpacs]{revtex4}

\usepackage{graphicx}
\usepackage{bm}

\newcommand{\avg}[1]{\langle #1 \rangle}
\newcommand{\avgs}[2]{\langle #1 \rangle({#2})}
\newcommand{\expv}[1]{{\mathrm{E}\!\left( #1 \right)}}
\newcommand{\expvc}[1]{{\mathrm{E_c}\!\left( #1 \right)}}
\newcommand{\pd}[1]{\frac{\partial}{\partial #1}}
\def\l{\lambda}

\begin{document}

\preprint{Bicocca-FT-01-30}
\title{On the growth of bounded trees} \author{Claudio Destri}
\email{Claudio.Destri@mib.infn.it} \author{Luca Donetti}
\email{Luca.Donetti@mib.infn.it} \affiliation{Dipartimento di Fisica
G. Occhialini, Universit\`a di Milano--Bicocca and INFN, Sezione di
Milano, Piazza delle Scienze 3 - I-20126 Milano, Italy}

\begin{abstract}
Bounded infinite graphs are defined on the basis of natural physical
requirements. When specialized to trees this definition leads to a
natural conjecture that the average connectivity dimension of bounded
trees cannot exceed two. We verify that this bound is saturated by a
class of random trees, in which case we derive also explicit
expressions for the growth probabilities.
\end{abstract}


\maketitle

\section{Introduction, summary and outlook}

Regular lattices are used in statistical mechanics and field theory to
describe crystal geometry or to discretize flat space--time. When
dealing with irregular structures that arise in the study of many
physical systems (for instance polymer gels, structural glasses or
fractals) as well as in the discretization of curved space--time, more
general structures are needed to describe the underlying geometry;
prototypes of such structures are graphs, discrete networks made of
nodes and links.

In particular we are interested in ``generic'', or even ``random''
realizations of infinite graphs, subject only to very general physical
requirements. Consider, as examples in condensed matter theory, the
infinite cluster of percolation theory or the fractal graphs of growth
aggregates; other examples in more abstract contexts such as
euclidean quantum gravity and random matrix theory, are graphs
modelizations of both space--time and ``target space''.

Two physical requirements common to these applications are: first, the
coordination of nodes is bounded (the number of nearest neighbors of
an atom, molecule or basic building block has a geometrical upper
bound); secondly, in the limit of large radius the surface must be
negligible with respect to volume (this is necessary if the structure
is embeddable as a whole in a finite--dimensional euclidean space).

Our main goal is the identification of suitable algorithms that can produce
graphs which do fulfill those requirements. Moreover, since these graphs
should be as ``generic'' as possible, that is with a minimal amount of
constraints, a third natural property we ask for is that of ``statistical
homogeneity''. By this we mean that the probability to find any finite
neighborhood should be the same around every node.

In this paper we restrict our attention to trees, that is graphs
without loops, for two main reasons. They are important on their own,
since they completely characterize special, physically relevant cases:
for example, the incipient infinite cluster at critical percolation is
essentially a tree \cite{percol}, while in two--dimensional quantum
gravity interacting with conformal matter, for certain values of the
central charge the metric collapses to that of branched polymers
\cite{jonsson}. Moreover, we want to use trees as a starting point for
building graph: our project is to generate graphs by adding links to
random trees and study properties such as the connectivity and
spectral dimension as the density of loops increase.

The outline of this paper is as follows. In next section we recall
some definitions of graph theory, define an infinite {\em bounded}
graph as a graph which satisfies the first two previously stated requirements,
and recall the correspondence between trees and branching
processes. Next, in section \ref{rtrees}, we give a constructive
definition of random trees by describing the algorithm used to build
them and discuss their statistical homogeneity. In section \ref{grfun}
we write the recursion for the growth probabilities and find by a
simple and original method their scaling form in the limit of large
sizes, from which we rederive the well known result that
their local connectivity (or intrinsic Hausdorff) dimension $d_{\rm
c}$ is equal to $2$. Using statistical homogeneity we then obtain the
same result for the {\em average} connectivity dimension ${\bar
d}_{\rm c}$.

Along this way the following conjecture appears very natural: the
average connectivity dimension for random trees is an upper bound for
all bounded trees; that is to say ${\bar d}_{\rm c} \le 2$ for all
bounded trees with saturation only in the case of random trees. There
are indeed many examples of bounded trees with local connectivity
dimension greater than two, for example $NT_D$ trees \cite{NTD} or
spanning trees of $D$--dimensional lattices with $D>2$, but in these
cases there are macroscopic inhomogeneities that cause ${\bar d}_{\rm
c}$ to be different from its local counterpart $d_{\rm c}$ and always
such that ${\bar d}_{\rm c} < 2$.  We think indeed that randomness and
(statistical) homogeneity maximize the average connectivity dimension.

\section{Basic definitions}
We consider only infinite connected graphs and for any such graph
$\mathcal G$ we call $G$ the set of its nodes (or sites, vertices). By
definition $G$ is in one--to--one correspondence with the set of
natural number and any specific choice of such a correspondence
defines an indexing, or labelling for $\mathcal G$. Unless otherwise
specified we shall consider the graph as unlabelled, in the sense that
we restrict our attention to properties that are independent on the
labelling.

The coordination (or degree) of a site $x\in G$ (that is the number of its
nearest neighbors) will be denoted by $z(x)$. The spherical shell $S(x,r)$
and the spherical ball $B(x,r)$ around any given site $x$ are defined as
\begin{gather*} 
	S(x,r)=\{y\in G: d(x,y) = r \} \\
	B(x,r)=\{y\in G: d(x,y)\le r \}
\end{gather*} 
where $d(x,y)$ is the chemical distance on the graph; we shall denote the
cardinality of these sets with $s(x,r)=|S(x,r)|$ (the surface area) and
$v(x,r)=|B(x,r)|$ (the volume) respectively. In particular we evidently
have $v(x,0)=1$, $v(x,1) = s(x,1)+1 = z(x)+1$ and in general
\begin{equation*}
	v(x,r+1) = v(x,r)+s(x,r+1)
\end{equation*}
The quantities $s(x,r)$ and $v(x,r)$ are example of
``good observables'', in the sense that they describe properties of the
unlabelled graph. More general observables of this type are the subsurfaces
$s_z(x,r)$ and subvolumes $v_z(x,r)$ counting nodes with fixed coordination
$z$ in the shell $S(x,r)$ and the ball $B(x,r)$.
 
The average of a function $F: G \to \mathbb{R}$ around the site $x$ is
defined as the infinite radius limit of the average on the ball with center
$x$, namely:
\begin{equation}\label{averagedef}
	\avgs{F}{x}=\lim_{r\to \infty} \frac{1}{v(x,r)}\sum_{y\in B(x,r)} F(y)
\end{equation}
Whenever $\avgs{F}{x}$ does not depend on $x$, we drop the relative
specification and regard it as the proper definition of graph--average
$\avg{F}$. Then the measure of a subset $A\subseteq G$ is identified with
the graph--average $\avg{\chi_A}$ of the characteristic function $\chi_A$
of the subset. In particular a central role is played by the the measures
of the subsets of nodes with coordination $z$, namely the {\em fraction} of
nodes with coordination $z$, which we denote by $f_z$; more precisely, we
should consider first the limit
\begin{equation}
	f_z(x)=\lim_{r\to \infty} \frac{v_z(x,r)}{v(x,r)}
\end{equation}
of the fractions within a given ball and then worry about the dependence on
$x$. We shall henceforth restrict out attention to $f-$graphs, that is
graphs such that $f_z(x)$ does not depend on $x$, or $f_z(x)=$ constant
$\equiv f_z$. By definition we have the normalization
\begin{equation}\label{f_norm}
	 \sum_z f_z = 1 
\end{equation}
Loosely speaking, one could say that $f_z$ represents the probability that
a node chosen ``at random'' has coordination $z$.

\subsection{Local and average connectivity dimensions}\label{cdim}
The quantities $v(x,r)$ introduced above are sometimes called {\em
local growth functions} of the graph. Their asymptotic behavior for
$r \to \infty$ gives the {\em (local) connectivity dimension} $d_{\rm c}$ 
\cite{suzuki} of
the graph (sometimes called also {\em intrinsic Hausdorff dimension})
\begin{equation*}
	d_{\rm c} = \lim_{r\to\infty} \frac{\log v(x,r)}{\log r}
\end{equation*}

The {\em average growth function} $\avg{v(r)}$ is the average of $v(x,r)$
over the graph and its asymptotic behavior is related to the {\em average
connectivity dimension} $\bar d_{\rm c}$ of the graph, defined as
\begin{equation*}
	\bar d_{\rm c} = \lim_{r\to\infty} \frac{\log\,\avg{v(x,r)}}{\log r}
\end{equation*}
Note that, even if it can be shown that $d_{\rm c}$ does not depend on
node $x$, $\bar d_{\rm c}$ can be different from $d_{\rm c}$ because
the limit $r \to \infty$ may not commute with the limit implied by the
definition of the average. These two parameters are known to coincide
on many ``regular'' graph (for example on $D$--dimensional lattices)
but are different on graphs which are manifestly inhomogeneous (as,
for example the comb--like\cite{comb} and $NT_D$ trees \cite{NTD}).

\subsection{Bounded graphs}
We define an infinite connected graph $\mathcal G$ as {\em bounded}
whenever, for any $x \in G$, the coordination is bounded
\begin{equation*}
	  z(x) \leq \bar z < +\infty
\end{equation*}
and the growth is volume--dominated, namely
\begin{equation}\label{boundgrowth} 
	 \lim_{r \to \infty}\frac{s(x,r)}{v(x,r)} = 0
\end{equation}
Actually, it is easy to realize that if the second condition holds for one
given node then by the first condition it holds for its nearest neighbors
and then, by recurrence, for any other site.  Moreover, these conditions
imply that the average of any bounded function does not depend on the
choice of the center of the ball \cite{drv1}. In particular, this
means that a bounded graph is a $f-$graph. 

\subsection{Trees and branching processes}\label{coords}
We shall from now on restrict our attention to {\em trees}, that is graphs
without any loop. For a given connected tree one can follow a recursive
procedure to find shells of increasing radius around any given site $x$:
\begin{itemize}
\item the first shell is formed by all nearest neighbors of $x$ and
therefore its size coincides with the coordination of $x$: 
\begin{equation} \label{rec_start}
	s(x,1) = z(x)
\end{equation}
\item for every node $y$ of coordination $z(y)$ in $S(x,r)$ there is one link 
 connecting $y$ to a node belonging to $S(x,r-1)$  and $z(y)-1$ 
 connecting it to nodes in $S(x,r+1)$; since there are no loops,
 all links to $S(x,r+1)$ are directed to distinct nodes, so that
\begin{equation}\label{rec_step}
	s(x,r+1)=\sum_{y\in S(x,r)} [z(y)-1]
\end{equation} 
\end{itemize}
Thus the tree defines a specific realization of a {\em branching process}
rooted at $x$, such that $x$ has $z(x)$ branches rooted on its nearest
neighbors while every other node $y$ on $S(x,r)$ has $z(y)-1$ branches
rooted on its nearest neighbors on $S(x,r+1)$ (notice that we may
equivalently consider all branches as rooted on links rather than
nodes). In turn this identifies a {\em coordination sequence}
$\{z_1,z_2,z_3,\ldots\}$, where $z_1=z(x)$ is the coordination of $x$,
$z_2,z_3,\ldots,z_{v(x,1)}$ are the coordinations of nodes on the first
shell, $z_{v(x,1)+1},z_{v(x,1)+2},\ldots,z_{v(x,2)}$ are the coordinations
of nodes on the second shell, and so on. The indexing on each shell is
uniquely fixed by the branch structure as soon as the branch roots on each
nodes are indexed (we assume in a consecutive manner).  Therefore a
coordination sequence identifies a unique labelled and rooted tree, while
there are many coordination sequences, obtained by permuting labelled
branches over any given node and by changing the root itself, that
correspond to the same unlabelled tree; the latter identifies therefore an
entire equivalence class of coordination sequences, or {\em isomorphic}
labelled trees.

\subsection{Bounded trees}
Given a branching process that builds an infinite rooted tree, from
equation \eqref{rec_step} one can obtain the rate of growth of the shells
\begin{equation*}
	s(x,r+1)-s(x,r) =\!\!\!\sum_{y\in S(x,r)} \!\!\![z(y)-2] = 
	\sum_{j=1}^{s(x,r)} (z_{v(x,r-1)+j} - 2)
\end{equation*} 
which, after integration and use of equation \eqref{rec_start}, gives
\begin{equation*}
	s(x,r+1)= 2 + \!\!\!\sum_{y\in B(x,r)} \!\!\![z(y)-2] = \, 
	2 + \sum_{j=1}^{v(x,r)} (z_j-2)
\end{equation*}
In the limit of infinite radius, if the tree is bounded according to equation
\eqref{boundgrowth}, the average coordination follows
\begin{equation*}
	\avg{z}-2 = \lim_{r\to\infty} \frac{s(x,r+1)}{v(x,r)} \le 
	(\bar z-1) \lim_{r\to\infty} \frac{s(x,r)}{v(x,r)} =0
\end{equation*}
We have dropped any reference to $x$ in $\avg{z}$ since this quantity does
not depend on it for a bounded graph. For this reason, by considering
shells centered around any other node $x'$ of the tree rooted at $x$, the
same average coordination $\avg{z}=2$ would have been obtained. Thus
$\avg{z}=2$ is a property of the unrooted tree. In fact, since a bounded
tree is an $f-$tree, it may equivalently be written as an average over the
``probabilities'' $f_z$:
\begin{equation}\label{f_avg}
	 \avg{z} = \sum_z z f_z = 2
\end{equation}

\subsection{Link orientation}
We have seen that a rooted tree has a natural orientation for all
links determined by the branching direction: the root $x$ has $z(x)$
outgoing links and any other site $y$ has $z(y)-1$ outgoing links and
one incoming links. Thus, with the unique exception of the root, there
is a one--to--one correspondence between nodes and incoming links, so
that $f_z$ is also the fraction of links leading to nodes with
coordination $z$; on the other hand the correspondence between nodes
and outgoing links is one to $z-1$ so that the fractions of links
departing from nodes with coordination $z$ is
\begin{equation}\label{tildef}
	 \tilde f_z=(z-1)f_z
\end{equation}
Note that the $\tilde f_z$ may be interpreted as properly normalized
probabilities on a bounded tree because of equations \eqref{f_norm} and
\eqref{f_avg}:
\begin{equation*}
 \sum_{z=1}^{\bar z} \tilde f_z = \sum_{z=1}^{\bar z} (z-1)f_z = 2-1 = 1
\end{equation*}

\section{Random trees}\label{rtrees}
There exist many different ways of defining and generating random
trees (random binary trees, pyramids of various order, linear
recursive trees, branched polymers and many others). We restrict our
attention to growth stochastic algorithms that are local and produce
infinite trees that are bounded and {\em statistically homogeneous}
(see below).

\subsection{Random branching processes}
Taking into account the correspondence between trees and equivalence
classes of coordination sequences $\{z_1,z_2,\ldots\}$ as described
above, the most immediate definition of such a random tree with
predefined coordination fractions $\{f_z,\; z=1,2,\ldots,\bar z\}$
fulfilling equations \eqref{f_norm} and \eqref{f_avg}, would be as a
random coordination sequence in which each $z_j$ is extracted from the
set $\{1,2,\ldots,\bar z\}$ with probability $f_{z_j}$. This is known
as the (critical) Galton--Watson stochastic branching process
\cite{brproc}: it satisfies the locality requirement that the
coordination of one node does not depend on that of any other node and
guarantees that the average coordination is two, but does not fulfill
our request that the tree grows indefinitely (the growth stops when
the last shell of the tree is made solely of node with coordination
1). In fact a classical problem of probability theory is the estimate
of the {\em survival probability} of such a branching process: one
finds that the probability that a tree has more than $N$ nodes to be
of order $N^{-1/2}$ \cite{brproc}. Another shortcoming of this
approach, as presented above, is the bias on the tree root due to fact
that it has as many branches as its coordination, while every other
node has one branch less than its coordination. This is easily
remedied by constraining the tree root to have coordination one; one
must then consider coordination sequences of the form
$\{z_1,z_2,z_3,\ldots\}$ with $z_1=1$. Then all nodes with index
greater than $1$ are statistically equivalent in the sense that: {\em
i}) each one of them is the root of a new random branching process
distributed exactly as the one rooted on node 1; {\em ii}) processes
rooted on different branches are statistically independent.

To overcame the extinction problem, one may consider an infinite
coordination sequence as a collection of disconnected finite trees of
any size (sometimes called {\em random forests} or {\em branched
polymers }), by simply using the first coordination extracted after
extinction as first branch root of new branching process. This
corresponds to a ``grand--canonical'' ensemble for the statistic of
random trees \cite{jonsson}.

We are looking instead for a process {\em preconditioned on
non--extinction} \cite{kesten} that builds one infinite connected
bounded tree ``at random''. Of course, for practical numerical
purposes, to build very large connected trees one needs only to make
several attempts with the standard Galton--Watson branching process
(with the root coordination $z(x)$ constrained to be one or extracted
with probability proportional to $f_z/z$) until the required size is
reached. Alternatively, to ensure that the tree grows indefinitely we
must implement in the branching process itself the condition of
non--extinction (this is our choice). The link orientability play here
a crucial role. Suppose in fact that an infinite connected tree is
already given and that we reconstruct it starting from a certain node
$x$ by extracting the right coordination sequence; $x$ is naturally
the root of the corresponding branching process, but since we assumed
the infinite tree to be there already, we may choose another node $o$
as root from which the link orientation propagates; for arbitrarily
large reconstructed balls $B(x,r)$, we may assume that $o$ does not
belong to $B(x,r)$, that is $r<d(x,o)$ (there is a set of measure one
of candidates $o$ which satisfy this condition on an infinite tree);
consider now, for any $r'=1,2,\ldots,r$, the $s(x,r')$ links between
$S(x,r')$ and $S(x,r'-1)$; one and only one of them is incoming, that
is oriented towards the inner shell, while the other $s(x,r')-1$ are
outgoing, that is oriented towards the outer shell; taking
equation \eqref{tildef} into account this is the essential observation to
define the stochastic building algorithm we are looking for; we call
it algorithm {\bf A}:
\begin{description}
\item{\bf A1}: extract the coordination $z_1$ for the root node $x$ with
probability $f_{z_1}$, then pick at random, with equal probability $1/z_1$,
which of the $z_1$ link on $x$ is the incoming one and mark it;
\item{\bf A2}: on the shell $S(x,r)$, for $r=1,2,3,\ldots$, extract the
coordination $z_{\rm M}$ of the node attached to the marked link with
probability $\tilde f_{z_{\rm M}}$, then pick at random, with
equal probability $(z_{\rm M}-1)^{-1}$, one of the newly added links and
move the mark on it; extract the coordinations $z_j$ of the other
$s(x,r)-1$ nodes on the shell with probability $f_{z_j}$.
\end{description}

We see that the marked links are exactly those forming the unique path
connecting $x$ to $o$. As this stochastic algorithm proceeds
indefinitely, $o$ is removed to infinity. Since the random choosing events
in the algorithm are all independent, the probability of a coordination
sequence corresponding to $B(x,r)$ reads therefore (here $v = v(x,r)$)
\begin{equation}\label{PB}
\begin{split}
	\text{Prob}\,[\{z_1,z_2,\ldots\,z_v\}] &= 
	\frac{s(x,r+1)}{s(x,1)} \!\!\!\prod_{y\in B(x,r)}\!\!\! 
	f_{z(y)} \\ &= \frac1{z_1} \bigg[2 + \sum_{j=1}^v (z_j-2)\bigg]
	\prod_{j=1}^v f_{z_j}
\end{split}
\end{equation}
where, except for the first shell, all factors of the form $(z_{\rm
M}-1)^{-1}$ have canceled against the factor $z_{\rm M}-1$ in $\tilde
f_{z_{\rm M}}$ (see equation \eqref{tildef}), and the factor $s(x,r+1)$
comes from the sum over all possible positions of the marked link on
the last shell $S(x,r)$ (this is indeed the probability for an
unmarked tree). We see that non--extinction is realized by forcing at
least one node per shell, the one which has the marked link as
incoming link, to have a coordination $z\ge 2$ extracted with the
modified probability \eqref{tildef}. This is done however in such a
way that no trace of the marks remains except that for the last
completed shell $S(x,r)$, which is eventually removed to infinity;
for finite $B(x,r)$ even this residual mark disappears by summing
over all its possible positions, as in expression~\eqref{PB}.

\subsection{Statistical homogeneity}\label{shomo}

The probability \eqref{PB} is clearly invariant under all permutations of
the coordinations that do not touch $z_1$. The different role of the
coordination $z_1=s(x,1)$ with respect to all the others is due to the
fact that coordination sequences correspond to rooted and labelled
trees. This asymmetry disappears when we consider the tree as unlabelled
and therefore multiply \eqref{PB} by the number of distinct labellings of
the rooted $B(x,r)$. This number is proportional, through a suitable
symmetry factor, to $z_1!\prod_{j>1}(z_j-1)!$ since the root has $z_1$
branches, while all other nodes have $z_j-1$ branches. By the same token, 
we could consider other algorithms for producing coordination sequences as
long as they provide the same probability for the unlabelled $B(x,r)$.
It is not difficult to realize that one such algorithm is the following
algorithm {\bf B}:
\begin{description}
\item{\bf B1}: extract the coordination $z_1$ for the root node $x$ with
probability $f_{z_1}$;
\item{\bf B2}: on the shell $S(x,r)$, for $r=1,2,3,\ldots$, extract the
first coordination $z=z_{v(x,r-1)+1}$ with 
probability $\tilde f_z$, then extract the coordinations $z_j$ of the other
$s(x,r)-1$ nodes on the shell with probability $f_{z_j}$.
\end{description}
With this algorithm a specific path is explicitly selected on the tree, but
this leaves no traces on the infinite random unlabelled tree.  In more
precise probabilistic terms the situation is described as follows:
\begin{itemize}
\item every instance of an infinite critical Galton--Watson branching process
conditioned on non--extinction has {\em almost surely} a {\em unique} path
$\Gamma$ (the {\em spine}) extending from the root to infinity
\cite{kesten}. 
\end{itemize}
Evidently the spine is just our path of marked links. As a
consequence, there exists the so--called {\em spinal decomposition} of the
infinite tree \cite{aldous}, which in our case, with the root coordination
$z_1$ extracted with probability $f_{z_1}$, may be stated as follows:
\begin{itemize}
\item 
the branching process that builds the tree may be regarded as a collection
of infinite independent subprocesses; one processes builds the spine
$\Gamma$ by extracting the coordinations of its nodes with probability
${\tilde f}_z$; each node $y\in\Gamma$ is the root of $z(y)-2$ ($z(x)-1$
if $y=x$ is the root) independent and identically distributed
critical Galton--Watson processes without condition on non--extinction;
\end{itemize}
We see therefore that if the tree is cut on every link of the spine
$\Gamma$, a Galton--Watson random forest is obtained. As discussed above,
this ``grand--canonical'' ensemble of random trees is statistically
homogeneous. It is also clear that the spine does not induce any
statistical inhomogeneity on the infinite connected random tree, since all
traces of the marks which identify the spine disappears as we have already
noticed in the calculation of the ball probability (see previous section);
moreover, as we shall see in the following, the spine is a zero measure set
in the tree.

\section{Growth on random trees}\label{grfun}
In this section we study the properties of local and average growth
functions on the infinite random trees introduced in the previous section.
Let us begin with the growth around the node $x$ used as root in the
building stochastic algorithm.
 
\subsection{Recursion rules}
From the explicit form \eqref{PB} of the probability we 
may quite easily derive recursion rules for the probability
distributions of the random variables  $v(x,r)$ and $s(x,r)$ (volume and 
surface area of the ball), that is
\begin{equation*}
	P_r(v,s) = \text{Prob}\,[\,v(x,r)=v\;\text{and}\;s(x,r)=s\,]
\end{equation*}
To denote the probability distribution for $v$ or $s$ only,
we define
\begin{gather*}
	P_r(v,\cdot) = \sum_s  P_r(v,s) \\
	P_r(\cdot\,,s) = \sum_v  P_r(v,s)
\end{gather*}
When $r=1$, from equation \eqref{PB} we read the probability of a ball
made of the root only, so that
\begin{equation*}
	P_1(v,s) = f_s\delta(v-s-1)
\end{equation*}
where $\delta(n)=1$ if $n=0$ and zero otherwise, while the use of the
iterative rule gives the Markovian recursion
\begin{equation}\label{rec_P}
	P_{r+1}(v',s')= \sum_{v,s} P_r(v,s)\,W_{s\to s'}\,\delta(v'-s'-v)
\end{equation}
with the surface--to--surface transition probabilities, easily derived from
\eqref{PB},
\begin{equation}\label{eq_W}
	W_{s\to s'} = \frac{s'}{s} \sum_{z_1,z_2,\ldots,z_s}
 	\delta \left(\sum_{i=1}^s (z_i-1)-s'\right) 
	\prod_{i=1}^s f_{z_i} 
\end{equation}
Now it is useful to introduce the generating functions, or discrete Laplace
transforms
\begin{equation*}
	 G_r(\l,\mu) = \sum_{v=1}^\infty 
	\sum_{s=1}^\infty P_r(v,s)\l^v \mu^s
\end{equation*}
for which the recursion reads
\begin{equation*}
 \begin{split}
	G_{r+1}(\l,\mu) & = \sum_{v'=1}^\infty 
	\sum_{s'=1}^\infty P_{r+1}(v',s')\l^{v'} \mu^{s'} \\
	& = \sum_{v=1}^\infty \sum_{s=1}^\infty P_r(v,s) \l^{v}
	\sum_{s'=1}^\infty W_{s\to s'}(\l\mu)^{s'} \\
	& = \sum_{v=1}^\infty \sum_{s=1}^\infty P_r(v,s) \l^{v}
	\l\mu\,g'(\l\mu)\, g(\l\mu)^{s-1}
 \end{split}
\end{equation*}
that is 
\begin{equation}\label{rec_G}
	G_{r+1}(\l,\mu)  = \l\mu \,\frac{g'(\l\mu)}{g(\l\mu)}\,
	G_r(\l,g(\l\mu))
\end{equation}
where
\begin{equation*}
 g(\l)  = \avg{\l^{z-1}} = \sum_{z=1}^{\bar z}  f_z \l^{z-1} =
  f_1 + f_2 \l +  f_3\l^2 + \ldots
\end{equation*}
is a polynomial in $\l$ with the properties
\begin{equation}\label{g_prop}
	 g(1)=1 \quad , \qquad g'(1)=1 
\end{equation}
which follow from equations \eqref{f_norm} and \eqref{f_avg}. At $r=1$
we have
\begin{equation*}
	G_1(\l,\mu) = \l^2\mu\, g(\l\mu)
\end{equation*}
while for $\l=\mu=1$ we verify
\begin{equation*}
	G_{r+1}(1,1) = G_r(1,1) = G_1(1,1) = 1
\end{equation*}
as required by the proper normalization for $P_r(v,s)$.  By
construction $G_r(1,\mu)$ is the discrete Laplace transform of
$P_r(\cdot\,,s)$ and satisfies the simpler recursion
\begin{equation}\label{rec_Gmu}
	G_{r+1}(1,\mu)  = \mu \,\frac{g'(\mu)}{g(\mu)}\,G_r(1,g(\mu))
\end{equation}
which may also be written as
\begin{equation}\label{rec_Gmug}
	G_{r+1}(1,\mu)  = \mu\, g_{r+1}(\mu)\, g'_r(\mu) \;,\quad
	g_{r+1}(\mu) = g(g_r(\mu)) \;,\quad g_0(\mu) = \mu
\end{equation}
Similarly $G_r(\l,1)$ is the discrete Laplace transform of $P_r(v,\cdot)$
and satisfies the recursion
\begin{equation}\label{rec_Gl}
	\frac{G_{r+1}(\l,1)}{h_{r+1}(\l)}  = 
	\l\,g'(h_{r-1}(\l)) \,\frac{G_r(\l,1)}{h_r(\l)}
\end{equation}
where $h_r(\l)$ fulfills its own independent recursion rule
\begin{equation}\label{rec_h}
	h_{r+1}(\l)  = \l\,g(h_r(\l))
\end{equation}
with the initial condition $h_0(\l)=\l$.

\begin{figure}
\includegraphics[height=100mm]{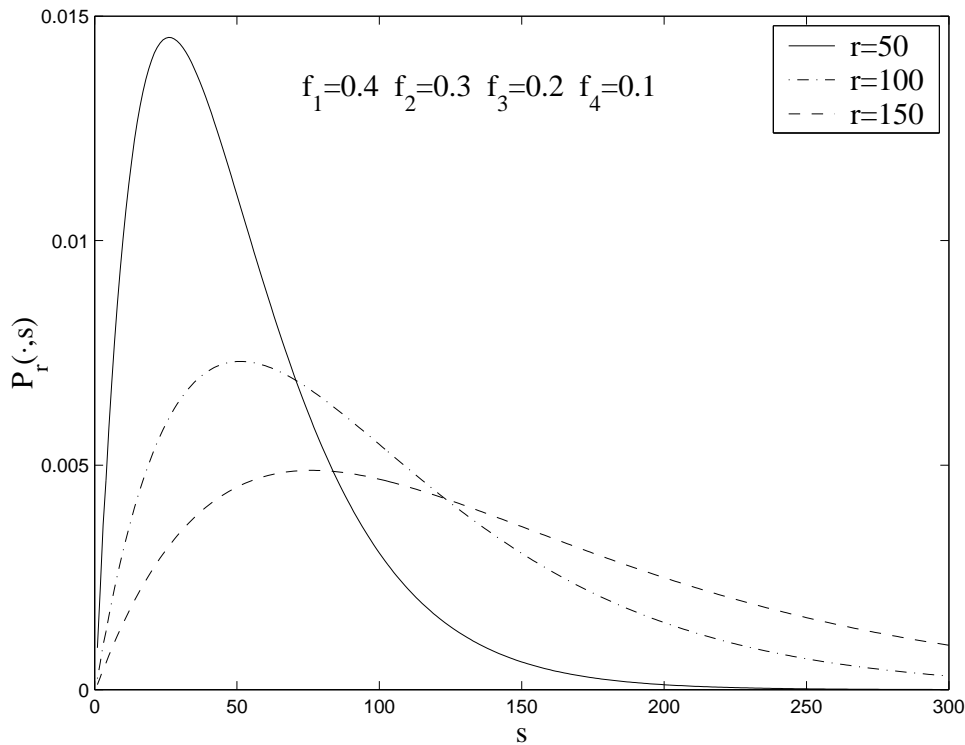}
\caption{\label{prs} Numerical evaluation of $P_r(\cdot\,,s)$}
\end{figure}

\begin{figure}
\includegraphics[height=100mm]{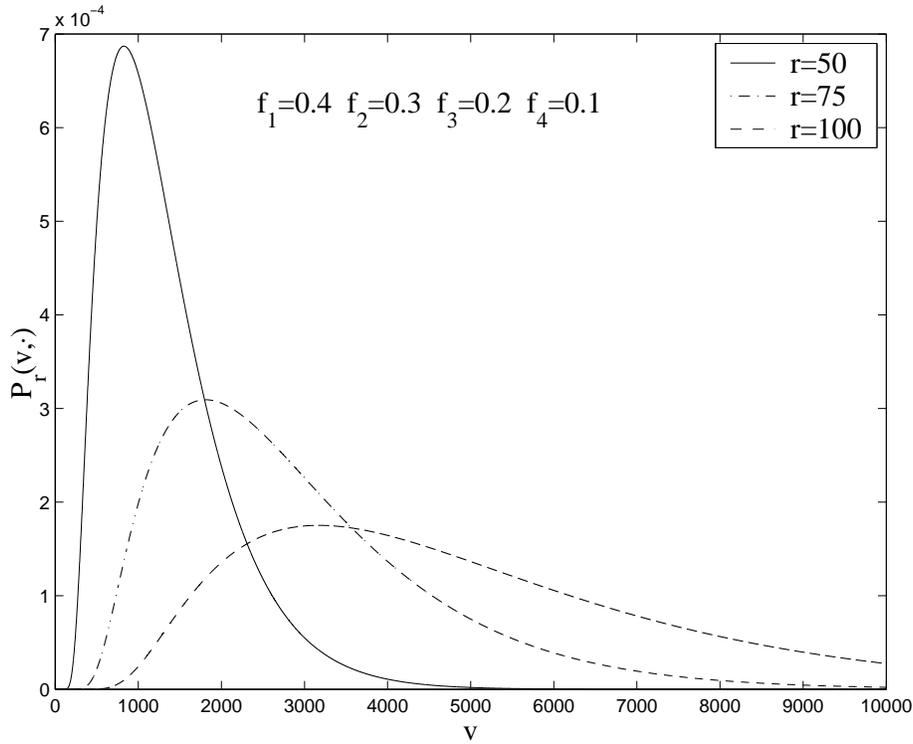}
\caption{\label{prv} Numerical evaluation of $P_r(v,\cdot)$}
\end{figure}

It should be stressed that the recursion rules \eqref{rec_G},
\eqref{rec_Gmu} and \eqref{rec_Gl} from a numerical point of view represent
already a complete solution of the original problem of determining the
probabilities $P_r(v,s)$, $P_r(\cdot\,,s)$ or $P_r(v,\cdot)$ for any $r$ (when
$r$ is small the problem is almost trivial for any symbolic manipulation
package). In fact we can choose $\l$ and $\mu$ to be the roots of unity of
order $N_1$ and $N_2$, respectively, with $N_1\le v_{\rm max}$ and $N_2\le
s_{\rm max}$, where $v_{\rm max}$, $s_{\rm max}$ are the largest values
at which $P_r(v,s)$ is nonzero. Then the recursion rules are used to
numerically find $G_r(\l,\mu)$ from the initial condition and finally
$P_r(v,s)$ is approximated by inverse discrete Fourier transform from
$G_r(\l,\mu)$ (of course, for large $r$ we must take $N_1$ and $N_2$ large
but fixed, much smaller than $v_{\rm max}$ and $s_{\rm max}$). We have
performed this program separately for surface and volume probabilities
using the numerical package {\sf Matlab} with the results plotted in
figures \ref{prs} and \ref{prv}.

        
\subsection{Expectation values and scaling}
The expectation values for the volume and surface area, defined as 
\begin{gather*}
	\expv{s(x,r)}=\sum_s s P_r(\cdot\,,s) = 
	\left. \pd{\mu} G_r(1,\mu) \right|_{\mu=1} \\
	\expv{v(x,r)}=\sum_v v P_r(v,\cdot) = 
	\left. \pd{\l} G_r(\l,1) \right|_{\l=1}
\end{gather*}
can be explicitly calculated from the recursion rules; in fact for the
first shell one simply gets
\begin{gather*}
	\expv{s(x,1)} = \avg{z} = 2\\
	\expv{v(x,1)} = 3
\end{gather*}
while for larger values of $r$ the recurrence relations are easily
found to read
\begin{equation*}
	\expv{s(x,r+1)} = \expv{s(x,r)} + 2\alpha
\end{equation*}
and, quite obviously
\begin{equation*}
	\expv{v(x,r+1)} = \expv{v(x,r)} + \expv{s(x,r)} + 2\alpha
\end{equation*}
where 
\begin{equation*}
	 2\alpha = g''(1) = \avg{z^2} - 4 = \avg{z^2} - \avg{z}^2 =
	(\Delta z)^2 
\end{equation*}
Thus we obtain 
\begin{equation}\label{expv_sv}
	 \expv{s(x,r)} = 2 + 2\alpha(r-1) \;,\quad
	 \expv{v(x,r)} = 1+2 r + \alpha (r-1)r
\end{equation}
By further differentiating equation \eqref{rec_G}, one finds recursion
rules for the expectation values of higher powers of $s(x,r)$ and $v(x,r)$
as well as of product of them (the moments of $P_r(v,s)$). By
differentiating the logarithm of \eqref{rec_G} one gets the centered, or
connected moments. For instance, the general form of the connected surface
moments is found to be
\begin{equation}
	\expvc{[s(x,r)]^k} = c_k \alpha^k r^k + O(r^{k-1})
\end{equation}
with $c_k$ constants; this fact suggests the scaling hypothesis that in the
limit $r \to \infty$ of the shell probability must be a function only of the
ratio $s/(\alpha r)$ (this is confirmed by the numerical determination in
figure \ref{prs}, after proper rescalings)
\begin{equation*}
	\alpha r P_r(\cdot\,,s)\ \overset{r \to \infty}{\sim} 
	\phi\Big(\frac{s}{\alpha r}\Big)
\end{equation*}
where the factor $\alpha r$ on the left comes from the integration
measure in the normalization:
\begin{equation}\label{Fnorm}
	1 =  \sum_{s=1}^{\infty} \frac1{\alpha r} 
	 \Big( \alpha r P_r(\cdot\,,s)\Big) \ 
	\overset{r \to\infty}{\sim} \int_0^\infty d\tau\, \phi(\tau)
\end{equation}
By the same token we see that the functions
\begin{equation*}
	F_r(u) = G_r \big(1,\exp(-\tfrac{u}{\alpha r}) \big)
	= \sum_{s=1}^{\infty} \frac1{\alpha r} 
	\Big(\alpha r P_r(\cdot\,,s)\Big)\exp(-\tfrac{us}{\alpha r})
\end{equation*}
should converge as $r\to\infty$ to the (continuous) Laplace transform
$F(u)$ of the the scaling function $\phi(\tau)$
\begin{equation*}
	F_r(u)\ \overset{r \to\infty}{\sim} \int_0^\infty d\tau\,
	 \phi(\tau)\, e^{-\tau u}  \equiv F(u) 
\end{equation*}
To calculate $F(u)$ we start from equation \eqref{rec_Gmug} and notice
that the property \eqref{g_prop} and the recursion $g_{r+1}(\mu) =
g(g_r(\mu))$ imply $g_r(1)=g'_r(1)=1$ for all $r$. This entails the
scaling law
\begin{equation*}
	g_r\big(\!\exp(-\tfrac{u}{\alpha r})\big) = 1 + 
	\frac{A(u)}{\alpha r}+ o\bigg(\frac1{r} \bigg)
\end{equation*}
with $A(0)=0$ and $A'(0)=-1$. Therefore
\begin{equation}\label{gtoF}
	F_r(u) = - \alpha\,r g_r\big(\!\exp(-\tfrac{u}{\alpha r})\big)\pd u 
	g_{r-1}\big(\!\exp(-\tfrac{u}{\alpha r})\big) \simeq -A'(u) + o(1)
\end{equation}
yielding the identification $F(u)=-A'(u)$. Then the recursion rule for
$g_r(\mu)$ to leading non--trivial order becomes the differential equation
for $A(u)$
\begin{equation*}
	uA' = A(1+A)
\end{equation*}
with the solution
\begin{equation*}
	A(u) = \frac{-u}{1+u}
\end{equation*}
Hence
\begin{equation}\label{Ffun}
	F(u)= \frac{1}{\left( 1+ u \right)^2}
\end{equation}
with the inverse Laplace transform
\begin{equation*}
	\phi(\tau) = \tau\, e^{-\tau}
\end{equation*}
so that the surface probability reads for large $r$
\begin{equation}
	P_r(\cdot\,,s) \sim \frac{s}{\alpha^2r^2}\, e^{-s/(\alpha r)}
\end{equation}
We may repeat this approach for the volume probabilities as well, by
assuming a scaling form in the variable $v/(\alpha r^2)$ (this would be obvious
from the surface scaling for an homogeneous structure and is confirmed
by the first connected moments and by the numerical data of
figure \ref{prv}).  Thus we set (notice that $g(1)=1$ implies $h_r(1)=1$,
see equation \eqref{rec_h})
\begin{equation*}
 	h_r\big(\!\exp(-\tfrac{\xi}{\alpha r^2})\big) 
	= 1 + \frac{B(\xi)}{\alpha r} + o\bigg(\frac1{r} \bigg)
\end{equation*}
Substituting this expression in equation \eqref{rec_h} we obtain to leading
non--trivial order
\begin{equation*}
	2 \xi B' = B(1+B) - \xi 
\end{equation*}
whose solution, taking care of the condition $B'(0)=-1$ that comes
from $h'_r(1)=r$, reads
\begin{equation*}
	B(\xi) = - \sqrt{\xi} \tanh(\sqrt{\xi})
\end{equation*}
Now we set
\begin{equation*}
  	G_r\big(\exp(-\tfrac{\xi}{\alpha r^2}),1\big) = H(\xi) + o(1)
\end{equation*}
so that
\begin{equation*}
 	\frac{G_{r+1}\big(\exp(-\tfrac{\xi}{\alpha r^2}),1\big)}
	{G_r\big(\exp(-\tfrac{\xi}{\alpha r^2}),1\big)} \simeq
	1 + \frac{2 \xi}{r} \frac{H'(\xi)}{H(\xi)}+ o\bigg(\frac1{r} \bigg)
\end{equation*}
and since for $\l=\exp(-\tfrac{\xi}{\alpha r^2})$
\begin{equation*}
	\l\,g'(\l h_r(\l)) \, \frac{h_{r+1}(\l)}{h_r(\l)} 
	 \simeq 1 + \frac2r B(\xi) + o\bigg(\frac1{r} \bigg)
\end{equation*}
substituting these expressions in equation \eqref{rec_Gl} we obtain
the equation for $H(\xi)$
\begin{equation*}
	\xi H'(\xi) = B(\xi) H(\xi)
\end{equation*}
and finally
\begin{equation}\label{Hfun}
	H(\xi) = \frac1{(\cosh\sqrt{\xi})^2}
\end{equation}
due to the normalization condition $H(0)=1$. The inverse Laplace transform 
of $H(\xi)$ is just the scaling form of the volume probabilities
\begin{equation*}
	 \alpha r^2 P_r(v,\cdot)\ \overset{r \to \infty}
	{\simeq} \Phi\Big(\frac{v}{\alpha r^2}\Big)
\end{equation*}
and may be written as (see Appendix \ref{invLapl} for details)
\begin{equation}\label{Pfun}
	\Phi(\tau) =  m^{1/4}\left(\frac{2K(m)}\pi \right)^{1/2}
	\left( \frac2\pi K(1-m)E(m) -1\right)
\end{equation}
where
\begin{equation*}
	\tau = \frac{K(1-m)}{\pi K(m)}
\end{equation*}
and $K(m)$, $E(m)$ are the two complete elliptic integrals
\begin{equation*}
	K(m) = \int_0^{\pi/2} \frac{d\theta}{\sqrt{1-m\sin^2\!\theta}} \;,
	\quad E(m) = \int_0^{\pi/2} d\theta \sqrt{1-m\sin^2\!\theta}
\end{equation*}
The large $\tau$ behavior of $\Phi(\tau)$ follows in the limit $m\to 0$,
\begin{equation*}
	m \simeq 16 \, e^{-\pi^2 \tau}
\end{equation*}
and reads
\begin{equation*}
	\Phi(\tau) \simeq (\pi^2 \tau -2)\, e^{-\pi^2 \tau/4}\;,\quad
	\tau\to\infty
\end{equation*}
The behavior for small $\tau$ is related to that for $m\to 1$,
\begin{equation*}
	1-m \simeq 16\, e^{-1/\tau}
\end{equation*}
so that 
\begin{equation*}
	\Phi(\tau) \simeq \frac4{\sqrt{\pi}}\, \tau^{-3/2}\,
	e^{-1/\tau}\;,\quad \tau\to 0
\end{equation*}
Therefore the volume probability distribution has the large $r$ asymptotic
behavior
\begin{equation}\label{main}
	P_r(v,\cdot) \simeq 
	\begin{cases}
	\dfrac{4\,\alpha^{1/2}r}{\pi^{1/2}v^{3/2}}\, e^{-\alpha r^2 /v} 
	& \text{for } v \ll r^2 \vspace{3mm}\\ 
	\vspace{2mm}
	\dfrac{\pi^2v}{\alpha^2 r^4}\,e^{-\pi^2 v /(4\alpha r^2)} 
	& \text{for } v \gg r^2 
	\end{cases}
\end{equation}

\subsection{Local and average connectivity dimensions}
The results of the previous section concern the probability distributions
for the growth around $x$ in the statistical ensemble of infinite trees
rooted at $x$. To relate these results to graph averages over a single
``generic'' realization of such a tree, we consider the following
observables
\begin{equation*}
	{\bar P}_r(v) = \frac1{v(x,R-r)} \sum_{y\in B(x,R-r)}
	\delta(v(y,r)-v)
\end{equation*}
that are the fractions of nodes $y$ for which the ball of radius $r$
has volume $v$ in a specific realization of a tree rooted at $x$
(restricted to radius $R$).  In the limit of $R\to\infty$ the natural
ergodic, or better {\em autoaveraging} hypothesis is that ${\bar
P}_r(v)$ deviates from its expectation value $\expv{{\bar P}_r(v)}$
only on a subset of measure zero of the statistical ensemble and that
$\expv{{\bar P}_r(v)}$ coincides with $P_r(v,\cdot)$, the previously
calculated probability
\begin{equation}\label{wlln}
	\lim_{R\to\infty} \Delta {\bar P}_r(v) = 0\;,\quad 
	\lim_{R\to\infty} \expv{{\bar P}_r(v)} = P_r(v,\cdot)
\end{equation}
As a matter of fact, the results reported in section \ref{shomo} are enough
to conclude that this autoaveraging is not simply an hypothesis: let us fix
the size $r$ of balls $B(y,r)$ centered on arbitrary nodes and consider the
local finite branching process rooted on $y$ that reconstructs $B(y,r)$; if
$y$ dists more than $r$ from from the spine this process is distributed
exactly like that rooted on $x$; if two such processes are rooted on
nodes well separated, they are independent; the number of nodes closer than
$r$ to the spine scales with $R$, the size of the main process, as compare
to the volume $v(x,R-r)$ that scales as $R^2$; from the (weak) law of large
numbers we then expect \eqref{wlln} to hold. We have performed several 
numerical checks that confirm this expectation with high accuracy (see
section \ref{numcheck}) 

We conclude therefore, by looking at the main results \eqref{main},
that the local connectivity (or intrinsic Hausdorff) dimension $d_{\rm
c} $ of section \ref{cdim} exists and that $d_{\rm c}=2$ with
probability one. Moreover, ${\bar d}_{\rm c} = d_{\rm c}$ because,
thanks to \eqref{wlln}, we can substitute the expectation value
$\expv{v(x,r)}$ in \eqref{expv_sv} with the graph average $\avg{v(r)}$
for a single random tree.

\subsection{Numerical checks}\label{numcheck}
\begin{figure}
\includegraphics[height=100mm]{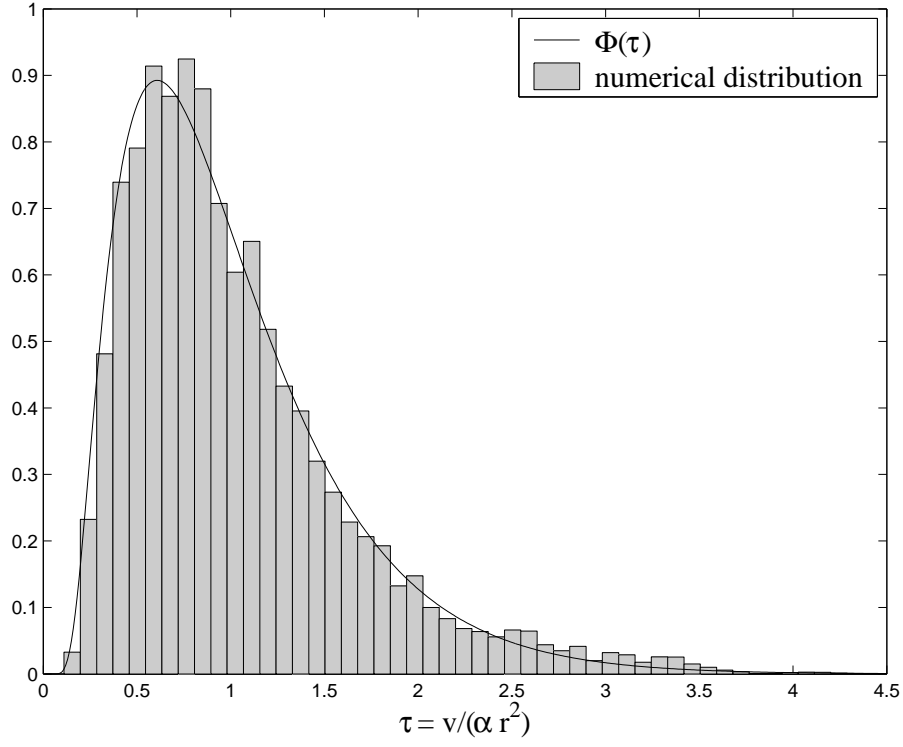}
\caption{\label{hist} Comparison between the distribution of
$v(y,r)/r^2$ ($r=500$) in a graph and $\Phi(\tau)$}
\end{figure}

We have performed several numerical checks of the autoaveraging
property discussed above; for example figure \ref{hist} compares the
distribution of $\tau=v(y,r)/(\alpha r^2)$ for $3\cdot 10^5$ nodes $y$
on a graph of $10^7$ nodes with $r=500$, with the calculated limit
distribution $\Phi(\tau)$.

\begin{figure}
\includegraphics[height=100mm]{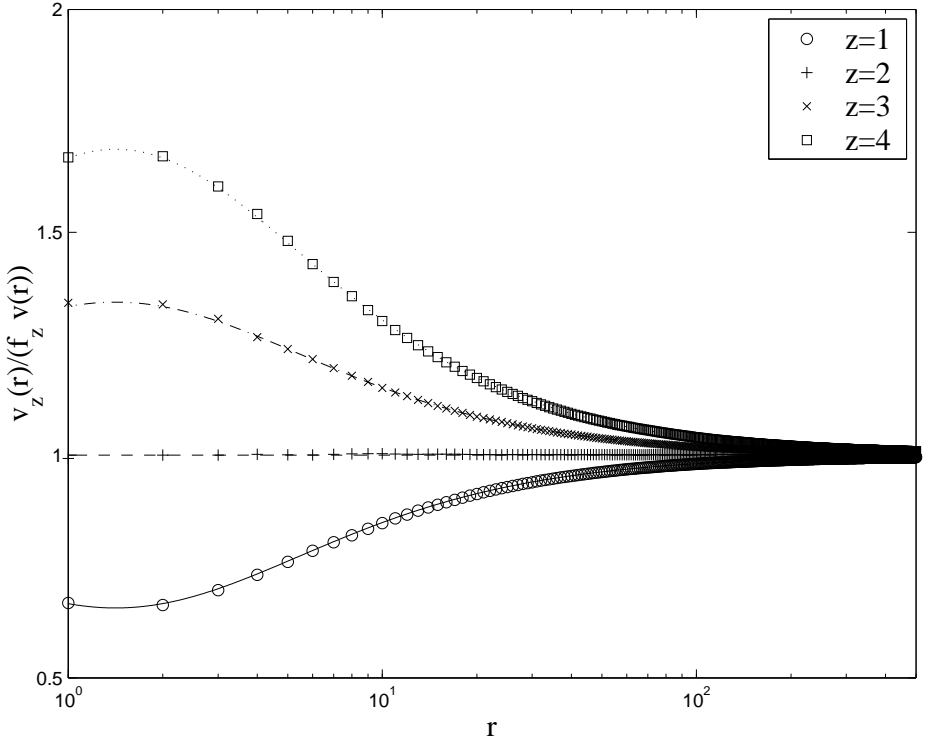}
\caption{\label{figvz} Comparison between numerical averages over
$3\cdot 10^4$ nodes of a single tree with $10^7$ and expectation
values. Error bars are smaller than the size of symbols}
\end{figure}

Another check can be done in the case of $v_z(x,r)$, since its average
over different realizations of the graph--generating algorithm can
easily be calculated and compared with the numerically obtained
average over different nodes on a single graph. The spinal
decomposition and the considerations in the previous section allow us
to write
\begin{equation*}
 \begin{split}
	\expv{v_z(x,r)} & =\tilde f_z r + (\expv{v(x,r)} - r) f_z  \\
	& = f_z \big(1 + (z-\alpha)r+\alpha r^2 \big)
 \end{split}  
\end{equation*}
with the use of equation \eqref{expv_sv}.
In figure \ref{figvz} we have plotted $\expv{v_z(x,r)}/(f_z \expv{v(x,r)})$
compared with the corresponding average over $3\cdot 10^4$ nodes of a single 
random tree with $10^7$ nodes.

\appendix

\section{Inverse Laplace transform of $H(\xi)$}
\label{invLapl}

The inverse Laplace transform of $H(\xi)$, $\Phi(\tau)$ is defined to be
\begin{equation*}
	\Phi(\tau) = \frac1{2 \pi i} \int_{\gamma-i\,\infty}^{\gamma+i\,\infty}
	d\xi H(\xi) \exp(\tau \xi) 
\end{equation*}
where $\gamma$ is an arbitrary positive constant chosen so that the contour 
of integration lies to the right of all singularities in $H(\xi)$.
In this case $H(\xi)$ is given by equation \eqref{Hfun}
\begin{equation*}
	H(\xi) = \frac1{(\cosh\sqrt{\xi})^2}
\end{equation*}
and its singularities in the complex plane of the variable $\xi$ are
located in
\begin{equation*}
	\xi_n = -\pi^2 \Big( n + \frac12 \Big)^2 \qquad n=0,1,2,\ldots
\end{equation*}
and the residue of $H(\xi) \exp(\tau \xi)$ in $\xi_n$ reads
\begin{equation*}
	\text{Res}[H(\xi) \exp(\tau \xi),\xi_n] = 
	\left(\pi^2 (2n+1)^2 \tau -2 \right) e^{-\pi^2 (n+1/2)^2 \tau}
\end{equation*}
Now using the residue theorem 
\begin{equation}\label{Phiseries}
 \begin{split} \Phi(\tau) & = \sum_{n=0}^{\infty}
	\text{Res}[H(\xi)\exp(\tau\xi),\xi_n] \\
	 & = \sum_{n=0}^{\infty} \left(\pi^2 (2n+1)^2 \tau -2 \right)
	e^{-\pi^2 (n+1/2)^2 \tau} \\
	 & = (-2 \tau \pd{\tau} -1)\ \theta_2(0,e^{-\pi^2 \tau})
 \end{split}
\end{equation}
where $\theta_2(u,q)$ is the second elliptic theta function with argument
$u$ and nome $q$; for $u=0$ it can be expressed in terms of the
complete elliptic integral $K(m)$
\begin{equation*}
	\theta_2(0,e^{-\pi^2 \tau}) =
	\left(\frac{2 m^{1/2}K(m)}\pi \right)^{1/2}
\end{equation*}
with $m$ obtained by inverting
\begin{equation*}
	\tau = \frac{K(1-m)}{\pi K(m)}
\end{equation*}
Using the identities
\begin{equation*}
	\pd{m}K(m)=\frac{E(m) - (1-m) K(m)}{2 m (1-m)}
\end{equation*}
and
\begin{equation*}
	K(m) E(1-m)+K(1-m) E(m) -K(m) K(1-m) = \frac{\pi}2
\end{equation*}
we obtain
\begin{equation*}
	-2\tau\pd{\tau} = \frac8\pi m(1-m)K(m)K(1-m) \pd{m}
\end{equation*}
and finally equation \eqref{Pfun}
\begin{equation*}
	\Phi(\tau) =  m^{1/4}\left(\frac{2K(m)}\pi \right)^{1/2}
	\left( \frac2\pi K(1-m)E(m) -1\right)
\end{equation*}
Now the asymptotic behavior for $\tau\to \infty$ and $\tau\to 0$ can be
found using the expansions
\begin{align*}
	K(m) \simeq &\frac\pi2 \big( 1+ \frac{m}4 \big) 
	  &&\text{for } m\to 0 \\
	E(m) \simeq &\frac\pi2 \big( 1- \frac{m}4 \big) 
	  &&\text{for } m\to 0 \\
	K(m) \simeq &\log4-\frac12\log(1-m) + \frac14 (1-m) 
	  \left(\log4-\frac12\log(1-m) -1 \right)
	  &&\text{for } m\to 1 \\
	E(m) \simeq &1+\frac14 (1-m)
	 \left(2\log4-\log(1-m) -1 \right)
	  &&\text{for } m\to 1
\end{align*}

For $m\to 0$ one can obtain
\begin{gather*}
	\tau \simeq \frac{4\log2 - \log m}{\pi^2} \\
	m \simeq 16\, e^{-\pi^2 \tau}
\end{gather*}
so that the asymptotic behavior $\Phi(\tau)$ for $\tau\to+\infty$
\begin{equation*}
 \begin{split}
	\Phi(\tau) & \simeq m^{1/4} \big(\log4 - \frac12 \log m -1 \big) \\
	 & \simeq (\pi^2 \tau -2)\, e^{-\pi^2 \tau/4} 
 \end{split}
\end{equation*}tar
that is simply the first term in the series equation \eqref{Phiseries}.

For $m\to 1$ the variable change reads
\begin{gather*}
	\tau \simeq \frac1{4\log2 - \log(1-m)} \\
	1-m \simeq 16\, e^{-1/\tau}
\end{gather*}
so that for $\tau\to 0$ 
\begin{equation*}
 \begin{split}
	\Phi(\tau) & \simeq \bigg( \frac{1-m}4 \bigg) 
	 \big(4\log2 - \log(1-m) \big)^{3/2} \\
	& \simeq \frac4{\sqrt{\pi}}\, \tau^{-3/2}\, e^{-1/\tau}
 \end{split}
\end{equation*}

\vskip 1cm
{\bf List of figures}
\begin{enumerate}
\item Numerical evaluation of $P_r(\cdot\,,s)$
\item Numerical evaluation of $P_r(v,\cdot)$ 
\item Comparison between the distribution of $v(y,r)/r^2$ ($r=500$) 
in a graph and $\Phi(\tau)$
\item Comparison between numerical averages over
$3\cdot 10^4$ nodes of a single tree with $10^7$ and expectation
values.
\end{enumerate}

\end{document}